\documentclass[11pt]{article}
\usepackage{epstopdf}
\usepackage{color}
\usepackage{subfigure}
\usepackage{amsmath}
\usepackage{amssymb}
\usepackage{graphicx,color}
\usepackage{cite}
\usepackage{enumerate}
\usepackage{amsthm}
\usepackage{amsfonts,mathrsfs}
\usepackage{geometry}
\usepackage{ifthen}

\begin{document}

\title{\bf Skew information-based coherence generating power of quantum channels}

\vskip0.1in
\author{\small Zhaoqi Wu$^{1}$, Lin Zhang$^{2}$\thanks{Corresponding author. E-mail:
godyalin@163.com}, Shao-Ming Fei$^{3,4}$\thanks{Corresponding
author. E-mail: feishm@cnu.edu.cn}, Jianhui Wang$^{5}$\\
{\small\it  1. Department of Mathematics, Nanchang University, Nanchang 330031, P R China} \\
{\small\it  2. Institute of Mathematics, Hangzhou Dianzi University, Hangzhou 310018, P R China}\\
{\small\it  3. School of Mathematical Sciences, Capital Normal University, Beijing 100048, P R China}\\
{\small\it  4. Max-Planck-Institute for Mathematics in the Sciences,
04103 Leipzig, Germany}\\
{\small\it  5. Department of Physics, Nanchang University, Nanchang
330031, P R China} }

\date{}
\maketitle

\noindent {\bf Abstract} {\small } We study the ability of a quantum
channel to generate quantum coherence when it applies to incoherent
states. We define the measure of coherence generating power (CGP)
for a generic quantum channel to be the average coherence generated
by the quantum channel acting on a uniform ensemble of incoherent
states based on the skew information-based coherence measure. We
present explicitly the analytical formulae of the CGP for any
arbitrary finite dimensional unitary channels. We derive the mean
value of the CGP over the unitary groups and investigate the
typicality of the normalized CGP. Furthermore, we give an upper
bound of the CGP for the convex combinations of unitary channels.
Detailed examples are provided to calculate exactly the values of
the CGP for the unitary channels related to specific quantum gates
and for some qubit channels.

\vskip 0.1 in



\vskip0.2in

\noindent {\bf 1. Introduction}

Quantum coherence is a distinctive feature of quantum systems
associated with the superposition principle. It plays pivotal roles
in quantum thermodynamics \cite{Horodecki}, quantum metrology
\cite{Marvian1} and quantum biology \cite{Lambert}. The
quantification of quantum coherence from a mathematical perspective,
however, has only been considered not long ago in \cite{Baumgratz},
where a rigorous framework for coherence measures has been proposed.
The past few years have witnessed a great interest in quantifying
quantum coherence by utilizing various distance measures such as
relative entropy, $l_1$-norm, intrinsic randomness, robustness of
coherence, max-relative entropy, fidelity, affinity, skew
information, generalized $\alpha$-$z$-relative R\'enyi entropy,
logarithmic coherence number, Schatten-$p$-norm and Fisher
information, etc.
\cite{Baumgratz,Yuan2015,Napoli2016,Bu1,Xiong1,Yuchangshui1,Wuzhaoqi1,Wuzhaoqi2,Luo1,Luo3,Pires,Zhuxuena,Xizhengjun,TongDM1,Bosyk,Liyongming,Yuchangshui2,Lilei}.
On the other hand, the problem of coherence distillation and
coherence dilution have also been discussed
\cite{Winter2016,Chitambar2016,Regula2018,Fang2018,Liu2019,Lami2019,Zhao2018},
and a complete theory of one-shot coherence distillation has been
formulated \cite{Zhao2019}. The average quantum coherence over the
pure state decompositions of a mixed quantum state has been
discussed in \cite{ZhaoMJ1}. Feasible methods have been introduced
to detect and estimate the coherence by constructing coherence
witnesses for any finite-dimensional states \cite{Zhangchengjie}.
Quantum coherence from other resource-theoretical perspectives such
as no-broadcasting of quantum coherence \cite{Lostaglio,Marvian2},
interconversion between quantum coherence and quantum entanglement
or quantum correlations
\cite{Streltsov,Chitambar2,ZhuHJ,MaJ,Luo5,KIM,WuKD2018} have also
been studied extensively. The study on coherence of quantum channels
has also attracted much attention
\cite{Hu2016,Dana2017,Korzekwa2018,Datta2018,Theurer2019,Xujianwei2019,Jin2021}.

The coherence measure defined in \cite{Baumgratz} is
basis-dependent. To get rid of the influence of the basis, the
average coherence with respect to mutually unbiased bases or all the
basis sets have been discussed \cite{Cheng,Luo4}. On the other hand,
random pure quantum states provide new perspectives for various
phenomena in quantum physics and quantum information processing
\cite{Collins}. Average coherence based on the relative entropy of
coherence and its typicality for random pure states and random mixed
states have been derived in \cite{Zhang2,Zhang3}, and the average
subentropy, coherence and entanglement of random mixed quantum
states have been discussed in \cite{Zhang4}.

The concepts of cohering power and de-cohering power of generic
quantum channels have been initially introduced by Mani and
Karimipour \cite{Mani}. By optimization on the output
coherence, the coherence generating power (CGP) of a quantum channel
has been defined to quantify the power of a channel in generating
quantum coherence. Many examples have been given to the qubit channels including
those induced by quantum gates. Different kinds of
operations which either preserve or generate coherence have also
been studied \cite{Misra,MGD}. It was Zanardi \emph{et al.}
\cite{Zanardi1,Zanardi2} who first utilize probabilistic averages to
study the CGP. By introducing a measure based on the average
coherence generated by the channel acting on a uniform ensemble of
incoherent states, a new method in quantifying the CGP of unitary
channels has been formulated. The coherence measure based on the
Hilbert-Schmidt norm has been exploited to derive explicitly the
analytical formulae of the CGP.

However, the Hilbert-Schmidt norm measure is not a well-defined
coherence measure since it does not possess the expected
monotonicity property. In \cite{Zhang6} by using the well-defined
relative entropy of coherence measure, Zhang \emph{et al.} has studied the quantification of
CGP for a generic quantum channel via probabilistic averages
and derived explicitly the analytical formulae of CGP for
unitary channels and deduced an upper bound for the CGP for unital
quantum channels. Since the skew information-based coherence is also a
well-defined coherence measure that can be experimentally measured,
it is of significance to calculate the CGP of a generic quantum
channel under the skew information-based coherence, instead of the
relative entropy of coherence. In this paper, we will solve this
problem.

The paper is arranged as follows. In Section 2, we first recall the
concepts of skew information and skew information-based coherence.
Then by adopting the probabilistic averages, we define the coherence
generating power of a generic quantum channel with respect to skew
information-based coherence. In Section 3, we present an explicit
analytical formula of the CGP via skew information-based coherence
for any unitary channels and calculate the CGP of unitary channels
induced by some specific quantum gates. Based on the formula given
in Section 3, we further compute the mean valued of the CGP and
discuss the typicality for the normalized CGP in Section 4. In
Section 5, we study the CGP for convex combinations of unitary
channels and derive the CGP for some important qubit channels.
Finally, we give some concluding remarks in Section 6.

\vskip0.1in

{\bf 2. CGP of quantum channels under skew information-based
coherence}

\vskip0.1in

Let $\mathcal{H}=\mathbb{C}^N$ be a Hilbert space of dimension $N$,
and $\mathrm{B}\mathcal{(H)}$, $\mathrm{S}\mathcal{(H)}$ and
$\mathrm{D}\mathcal{(H)}$ be the set of all bounded linear
operators, Hermitian operators and density operators on
$\mathcal{H}$, respectively. Denote by $\mathrm{U(N)}$ the group of
all $N\times N$ unitary matrices.

Fix an orthonormal basis $\{|k\rangle\}^N_{k=1}$ of $\mathcal{H}$.
The set of incoherent states, which are diagonal in this basis, can
be written as $\mathcal{I}=\{\delta\in
\mathrm{D}\mathcal{(H)}|\delta=\sum^N_{k=1}p_k|k\rangle\langle
k|,~p_k\geq 0,~\sum^N_{k}p_k=1\}$. Let $\Lambda$ be a CPTP map
$\Lambda(\rho)=\sum_{n}K_n\rho K_n^\dag,$ where $K_n$ are Kraus
operators satisfying $\sum_{n}K_n^\dag K_n=I_{N}$ with $I_N$ the
identity operator on $\mathcal{H}$. $K_n$ are called incoherent
Kraus operators if $K_n^\dag \mathcal{I}K_n\in \mathcal{I}$ for all
$n$, and the corresponding $\Lambda$ is called an incoherent
operation.

A well-defined coherence measure $C(\cdot)$ of a quantum state
$\rho$ should satisfy the following conditions \cite{Baumgratz}:
\begin{itemize}
\item $(C1)$ (Faithfulness) $C(\rho)\geq 0$ and $C(\rho)=0$ iff $\rho$ is
incoherent;
\item $(C2)$ (Convexity) $C(\cdot)$ is convex in $\rho$;
\item $(C3)$ (Monotonicity) $C(\Lambda(\rho))\leq C(\rho)$ for any
incoherent operation $\Lambda$;
\item $(C4)$ (Strong monotonicity)
$C(\cdot)$ does not increase on average under selective incoherent
operations, i.e., $C(\rho)\geq \sum_{n}p_nC(\varrho_n),$ where
$p_n=\mathrm{Tr}(K_n\rho K_n^\dag)$ are probabilities and
$\varrho_n=\frac{K_n\rho K_n^\dag}{p_n}$ are the post-measurement
states, $K_n$ are incoherent Kraus operators.
\end{itemize}

The skew information-based coherence $C_S(\rho)$ of a quantum state $\rho$
with respect to a fixed orthonormal basis $\{|k\rangle\}^N_{i=1}$ in an
$N$-dimensional Hilbert space $H$ is defined by \cite{Yuchangshui1},
\begin{equation}\label{eq1}
C_S(\rho)=\sum_{k=1}^N I(\rho,|k\rangle\langle k|)=1-\sum_{k=1}^N \langle k|\sqrt{\rho}|k\rangle ^2,
\end{equation}
where $I(\rho,|k\rangle\langle
k|)=-\frac{1}{2}\mathrm{Tr}\{[\sqrt{\rho},|k\rangle \langle k|]\}^2$
is the skew information of the state $\rho$ with respect to the
projector $|k\rangle \langle k|$, $k=1,2,\cdots,N$. $C_S(\rho)$ is
shown to be a well-defined coherence measure which satisfies the
required properties of a coherence measure in the framework of
\cite{Baumgratz}. It is of pivotal importance with meaningful
physical interpretations and can be experimentally implemented. The
advantage of this coherence measure is that it has an analytic
expression. Also, an operational meaning in connection with quantum
metrology has been revealed. The distribution of this coherence
measure among the multipartite systems has been investigated and a
corresponding polygamy relation has been proposed. It is also found
that this coherence measure provides the natural upper bounds of
quantum correlations prepared by incoherent operations. Moreover, it
is shown that this coherence measure can be experimentally measured
\cite{Yuchangshui1}. Since the skew information-based coherence
measure (\ref{eq1}) is well-defined and can be analytically
expressed, it is of great significance both theoretically and
practically, and worth evaluating the CGP of unitary channels based
on this measure. Note that $C_S(\rho)$ attains the maximal value
$1-\frac{1}{N}$ at the maximal coherent sate
$|\psi\rangle=\frac{1}{\sqrt{N}}\sum_{j=1}^N
e^{i\theta_j}|j\rangle$.

Quantum ensembles are formulated by specifying probability measures
on $\mathrm{D}(\mathbb{C}^N)$. The uniqueness for such measures
cannot be guaranteed, while the Fubini-Study (FS) measure is the
only natural measure in defining random pure states
\cite{Bengtsson}.

Conventionally, it is not easy to deal with the emerged monotone
metrics when $N>2$, we have to take great efforts when we consider a
Riemannian geometry on $\mathrm{D}(\mathbb{C}^N)$. However, for some
special monotone metrics, the measures induced from them would be
easier to tackle with. Recall that in flat space, the Euclidean
measure is decomposed into a product. We can use the same technique
here. The set of quantum mixed states in the form $\rho=U\Lambda
U^\dagger$, with $\Lambda$ a fixed diagonal matrix having strictly
positive eigenvalues, is a flag manifold ${\bf
F}^{(N)}=\mathrm{U(N)/[U(1)]}^N$. If the chosen eigenvalues and
eigenvectors are independent, and the eigenvectors are drawn
according to the invariant Haar measure,
$\mathrm{d\mu_{\mathrm{Haar}}}(W)=\mathrm{d\mu_{\mathrm{Haar}}}(UW)$,
then we can assume that a probability distribution in
$\mathrm{D}(\mathbb{C}^N)$ possess the invariance with respect to
unitary rotations, $P(\rho)=P(W\rho W^\dagger)$\cite{Bengtsson}.

Combining the two measures, a product measure on the Cartesian
product of the flag manifold and the simplex ${\bf F}^{(N)}\times
\Delta_{N-1}$ can be defined: $\mathrm{d\omega(\rho)=
d\mu_{Haar}}(U)\times \mathrm{d}\mu(\Lambda)$, which induces the
corresponding probability distribution,
$P(\rho)=P_{\mathrm{Haar}}({\bf F}^{(N)})\times P(\Lambda)$, where
the first factor denotes the natural, unitarily invariant
distribution on the flag manifold ${\bf
F}^{(N)}=\mathrm{U(N)/[U(1)]}^N$ induced by the Haar measure on
$\mathrm{U(N)}$. Note that the Haar measure on $\mathrm{U(N)}$ is
unique while there is no unique choice for $\mu$ \cite{Bengtsson}.

The measures used frequently over $\mathrm{D}(\mathbb{C}^N)$ can be
obtained by taking partial trace over a $M$-dimensional environment
of an ensemble of pure states distributed according to the unique,
unitarily invariant FS measure on the space
$\mathbb{C}\mathrm{P}^{MN-1}$ of pure states of the composite
system. There is a simple physical motivation for such measures:
they can be used if anything is known about the density matrix,
apart from the dimensionality $M$ of the environment. When $M=1$, we
get the FS measure on the space of pure states. Since the rank of
$\rho$ is limited by $M$, when $M\geq N$ the induced measure covers
the full set of $\mathrm{D}(\mathbb{C}^N)$. Since the pure state
$|\psi\rangle$ is drawn according to the FS measure, the induced
measure is of the product form $P(\rho)=P_{\mathrm{Haar}}({\bf
F}^{(N)})\times P(\Lambda)$. Hence the distribution of the
eigenvectors of $\rho$ is determined by the Haar measure on
$\mathrm{U(N)}$ \cite{Bengtsson}.

The general measure for the joint probability distribution of
spectrum $\Lambda=\{\lambda_1,\ldots,\lambda_N\}$ of $\rho$ is given
by \cite{Zyczkowski},
\begin{eqnarray}\label{eq2}
\mathrm{d}\omega_{N,M} (\Lambda)=
C_{N,M}\delta\left(1-\sum^N_{j=1}\lambda_j\right)\prod_{1\leq
i<j\leq
N}(\lambda_i-\lambda_j)^2\prod^N_{j=1}\lambda^{M-N}_j\theta(\lambda_j)\mathrm{d}\lambda_j,
\end{eqnarray}
where $\delta$ is the Dirac delta function, the theta function
$\theta$ ensures that $\rho$ is positive definite, and $C_{N,M}$ is
the normalization constant,
$$
C_{N,M}=
\frac{\Gamma(NM)}{\prod^{N-1}_{j=0}\Gamma(N-j+1)\Gamma(M-j)}.
$$
In this paper we take $N=M$. In this scenario, we deal with
non-Hermitian square random matrice characteristic of the {\it
Ginibre ensemble} \cite{Ginibre,Mehta} and obtain the
Hilbert-Schmidt measure \cite{Bengtsson}. Denote
$\mathrm{d\omega_{N,N}=d\omega_{HS}}$ and
$C_{N,N}=C_N^{\mathrm{HS}}$. Thus we have \cite{Zhang5,Zyczkowski}
$$
\mathrm{d\omega_{HS}(\rho)=d\mu_{Haar}}(U)\times
\mathrm{d}\mu(\Lambda)
$$
for $\rho=U\Lambda U^\dagger$. Here $\mathrm{d\mu(\Lambda)}$ is
given by \cite{Zhang5,Zyczkowski},
\begin{equation}\label{eq3}
\mathrm{d\mu(\Lambda)}=C_N^{\mathrm{HS}}\delta\left(1-\sum_{j=1}^N\lambda_j\right)|\Delta(\lambda)|^2\prod_{j=1}^N
\mathrm{d}\lambda_j,
\end{equation}
where $\Delta(\lambda)=\prod_{1\leq k<l\leq N}(\lambda_l-\lambda_k)$
and
\begin{equation}\label{eq4}
C_N^{\mathrm{HS}}=\frac{\Gamma(N^2)}{\Gamma(N+1)\prod^{N}_{j=1}\Gamma(j)^2}.
\end{equation}

Let $\Phi$ be a quantum channel, i.e., a trace-preserving completely
positive linear map, which maps an incoherent state $\Lambda$ to
$\Phi(\Lambda)$. By employing the technique of probabilistic
averages \cite{Zanardi1,Zanardi2,Zhang6}, we define the measure of
coherence generating power (CGP) $\mathrm{\bf CGP_S}(\Phi)$ of
$\Phi$ to be the average skew information-based coherence generated
by the quantum channel acting on a uniform ensemble of incoherent
states,
\begin{equation}\label{eq5}
\mathrm{\bf
CGP_S}(\Phi):=\int_{\textcolor{blue}{\mathcal{I}}}\mathrm{d}\mu(\Lambda)C_S(\Phi(\Lambda)),
\end{equation}
where $\mathcal{I}$ denotes the set of incoherent states, $\mu$ is
the probability measure on a uniform ensemble of incoherent states
and $\mathrm{d}\mu(\Lambda)$ is given in Eq. (\ref{eq3}). Obviously,
for incoherent quantum channels $\Phi_{IO}$, one has $\mathrm{\bf
CGP_S}(\Phi_{IO})=0$ since $\Phi_{IO}(\Lambda)$ is always
incoherent.

In the following we calculate $\mathrm{\bf CGP_S}(\Phi)$ for unitary
channels $\Phi_{U}$ such that $\Phi_{U}(\Lambda)=U\Lambda U^{\dag}$,
where $U$ denotes unitary transformations and $\dag$ the transpose
and conjugation.

\vskip0.1in

\noindent {\bf 3. CGP of unitary channels under skew
information-based coherence}

\vskip0.1in

We first calculate the CGP of unitary channels under skew information-based coherence, see proof in Appendix A.

{\bf Theorem 3.1} For any given $N\times N$ unitary matrix $U$, the
$\mathrm{CGP}$ of the unitary channel $\Phi_{U}$ is given by
\begin{eqnarray}\label{eq6}
&&\mathrm{\bf CGP_S}(U):=\mathrm{\bf CGP_S}(\Phi_U)\nonumber\\
&&=\left(1-\frac{1}{N^2(N-1)}\left[\left(\sum_{k=1}^N
I_{kk}^{(\frac{1}{2})}\right)^2-\sum_{k,l=1}^N
\left(I_{kl}^{(\frac{1}{2})}\right)^2\right]\right)\left(1-\frac{1}{N}\sum_{k,i=1}^N|U_{ki}|^4\right),
\end{eqnarray}
where $
I_{kl}^{(\frac{1}{2})}=\sum_{r=0}^{\min(k,l)}(-1)^{k+l}\tbinom{\frac{1}{2}}{k-r}\tbinom{\frac{1}{2}}{l-r}\frac{\Gamma(\frac{3}{2}+r)}{r!}
$.

Below we present an estimation on the lower and upper bounds on $\mathrm{\bf CGP_S}(U)$.

{\bf Proposition 3.1} For any unitary channel $\Phi_U$,
$$
0\leq \mathrm{\bf CGP_S}(U)\leq \mathrm{CGP}_N,
$$
where
\begin{eqnarray}\label{eq7}
\mathrm{CGP}_N:=\left(1-\frac{1}{N}\right)\left(1-\frac{1}{N^2(N-1)}\left[\left(\sum_{k=1}^N
I_{kk}^{(\frac{1}{2})}\right)^2-\sum_{k,l=1}^N
\left(I_{kl}^{(\frac{1}{2})}\right)^2\right]\right).
\end{eqnarray}
The lower bound is saturated iff $|U_{ki}|\cdot|U_{kj}|=0$ for all
$k,i,j=1,2,\cdots,N$ with $i\neq j$, while the upper bound is
saturated iff $|U_{ki}|^2=1/N$ for all $k,i=1,2,\cdots,N$.

{\bf Proof.} Since $U$ is unitary, we have
$\sum^N_{i=1}|U_{ki}|^2=1$ for $k=1,2,\cdots,N$. Then
\begin{eqnarray*}
\sum_{k,i=1}^N|U_{ki}|^4=\sum_{k=1}^N\left(\sum_{i=1}^N
(|U_{ki}|^2)^2\right)\leq \sum_{k=1}^N\left(\sum_{i=1}^N
|U_{ki}|^2\right)^2=N.
\end{eqnarray*}
By Eq. (\ref{eq6}), we obtain that $\mathrm{\bf CGP_S}(U)\geq 0$. It
is easy to see that the lower bound is saturated, i.e., $\mathrm{\bf
CGP_S}(U)=0$, iff $\sum_{k,i=1}^N|U_{ki}|^4=N$ iff $\sum_{i=1}^N
(|U_{ki}|^2)^2= \left(\sum_{i=1}^N |U_{ki}|^2\right)^2$ iff
$|U_{ki}|\cdot|U_{kj}|=0$ for all $k,i,j=1,2,\cdots,N$ with $i\neq
j$.

On the other hand, noting that $\sum^N_{k,i=1}|U_{ki}|^2=N$ for
$k,i=1,2,\cdots,N$, and utilizing the Lagrange multiplier method,
one can check that the minimal value of $\sum^N_{k,i=1}|U_{ki}|^4$
is $1$, which is attained iff $|U_{ki}|^2=1/N$ for all
$k,i=1,2,\cdots,N$. This implies from (\ref{eq6}) that $\mathrm{\bf
CGP_S}(U)\leq \mathrm{CGP}_{N}$, and the upper bound is saturated
iff $|U_{ki}|^2=1/N$ for all $k,i=1,2,\cdots,N$. $\Box$

Note that $|U_{ki}|\cdot|U_{kj}|=0$ for all $k,i,j=1,2,\cdots,N$
with $i\neq j$ implies that at least one of the elements in each row
of the matrix is $0$. For example, when $N=2$, the unitary $U\in
\mathrm{U(2)}$ is in the following form,
\begin{equation*}
\left({\begin{array}{cc}
                    u & -v \\
                    \bar{v} & \bar{u}
\end{array}}\right),\ \ \
u,v\in \mathbb{C},\ \ \,|u|^2+|v|^2=1.
\end{equation*}
In order for the $\mathrm{\bf CGP_S}(U)$ to reach the lower bound
$0$, the unitary $U$ is in either of the following forms,
\begin{equation*}
\left({\begin{array}{cc}
                        0 & -e^{\sqrt{-1}\phi} \\
                        e^{-\sqrt{-1}\phi} & 0
\end{array}}\right)\ \ \ \makebox{or}\ \ \
\left({\begin{array}{cc}
                        e^{\sqrt{-1}\varphi} & 0 \\
                        0 & e^{-\sqrt{-1}\varphi}
\end{array}}\right).
\end{equation*}

A set of orthonormal bases $\{e_k\}$ with
$e_k=\{|0\rangle_k,|1\rangle_k,\cdots,|N-1\rangle_k\}$ for a Hilbert
space $H=\mathbb{C}^N$ is called mutually unbiased bases (MUBs) if
\cite{JSchwinger,IDIvanovic} $|_k\langle i|j\rangle_l|=1/\sqrt{N}$
holds for all $i,j\in \{0,1,\cdots,N-1\}$ and $k\neq l$. From
Proposition 3.1, it can be seen that if the base
$\{|i\rangle\}^N_{i=1}$ and the base $\{U|i\rangle\}^N_{i=1}$ are
mutually unbiased, the unitary channel $\Phi_U$ reaches the maximal
value of CGP. For example, the unitary $U$ satisfying that $\langle
s|U|t\rangle=1/\sqrt{N}\exp(\sqrt{-1}\frac{2\pi}{N}st)(s,t=1,2,\cdots,N)$
has the maximal CGP.

\begin{figure}[htbp]\centering
\includegraphics[width=0.5\textwidth]{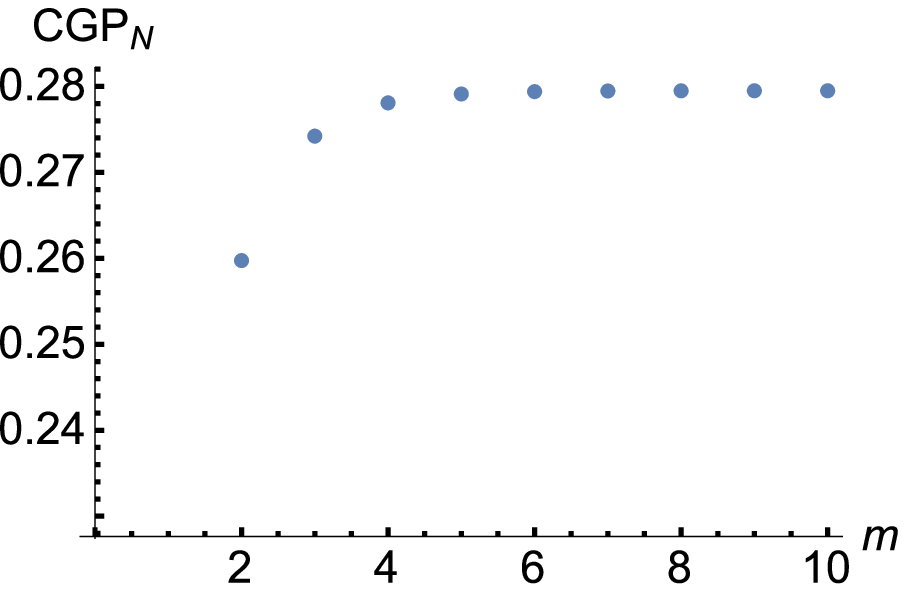}
\caption{The upper bound $\mathrm{CGP}_N$ of $\mathrm{\bf CGP_S}(U)$
as a function of $N=2^m$.} \label{fig:Fig1}
\end{figure}

From (\ref{eq7}), for $N=2$ and $N=3$ we have
$$\mathrm{CGP}_2=\frac{1}{2}\left(1-\frac{3\pi}{16}\right)\approx 0.205$$
and
$$\mathrm{CGP}_3=\frac{2}{3}\left(1-\frac{103\pi}{512}\right)\approx 0.245,$$
respectively. In Fig.~\ref{fig:Fig1} we plot the maximal value $\mathrm{CGP}_N$ of $\mathrm{\bf
CGP_S}(U)$ as a function of $N=2^m$ for $N=2,...,10$.
It shows that as $N$ increases, $\mathrm{CGP}_N$ approaches to $0.28$.

Next, as examples we calculate the CGP for some specific unitary
channels by using Theorem 3.1. First, for $N=2$ it follows from Eq.
(\ref{eq6}) that
\begin{eqnarray}\label{eq8}
\mathrm{\bf CGP_S}(U)=\left(1-\frac{3\pi}{16}\right)\left(1-
\frac{1}{2}\sum_{k,i=1}^2|U_{ki}|^4\right).
\end{eqnarray}

{\bf Example 3.1} Consider the Hadamard gate
$H=\frac1{\sqrt{2}}\left({\begin{array}{cc}
                         1 & 1 \\
                         1 & -1
                   \end{array}}\right)$.
From Eq. (\ref{eq8}) we have $\mathrm{\bf
CGP_S}(H)=\frac{1}{2}(1-\frac{3\pi}{16})\approx 0.205$.

{\bf Example 3.2} Consider the unitary transformation
$U_{\theta}=\left({\begin{array}{cc}
                             \cos\theta & \sin\theta \\
                             -\sin\theta & \cos\theta
                             \end{array}}\right)$.
By (\ref{eq8}), we have the CGP of the unitary channel related to
$U_{\theta}$,
\begin{equation}\label{eq9}
\mathrm{\bf CGP_S}(U_{\theta})=\frac{1}{2}\left(1-\frac{3\pi}{16}
\right)\sin^2{2\theta}.
\end{equation}
Fig. 2 shows the coherence generating power $\mathrm{\bf CGP_S}(U_\theta)$ of $U_{\theta}$
as the function of $\theta\in[0,\pi]$. It can be seen that the
maximal value of $\mathrm{\bf CGP_S}(U_\theta)$ is
$\frac{1}{2}(1-\frac{3\pi}{16})\approx 0.205$, which is attained at
$\theta=\pi/4$ and $\theta=3\pi/4$.
\begin{figure}[htbp]\centering
\includegraphics[width=0.5\textwidth]{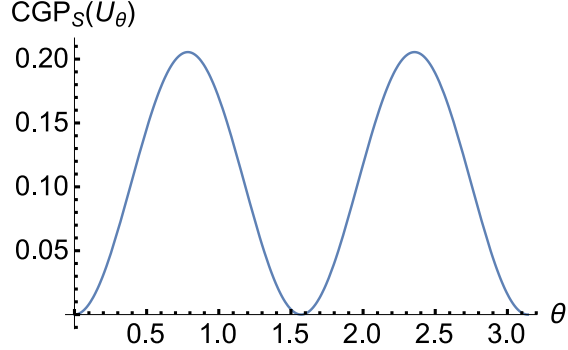}
\caption{The skew information-based coherence generating power of
$U_{\theta}$ with respect to the parameter $\theta$.}
\label{fig:Fig2}
\end{figure}

When $N=4$, it follows from Eq. (\ref{eq6}) that
\begin{eqnarray}\label{eq10}
\mathrm{\bf CGP_S}(U)=\left(1-\frac{54545\pi}{262144}\right)\left(1-
\frac{1}{4}\sum_{k,i=1}^4|U_{ki}|^4\right).
\end{eqnarray}

{\bf Example 3.3} (Square root of swap gate) The
$\sqrt{\text{swap}}$ gate is an important quantum gate since any quantum
multi-qubit gates can be generated by combining $\sqrt{\text{swap}}$ and single qubit gates, which is given by
$$
\sqrt{\text{swap}} =\left({\begin{array}{cccc}
                           1 & 0 & 0 & 0 \\
                           0 & \frac12(1+\mathrm{i}) & \frac12(1-\mathrm{i}) & 0 \\
                           0 & \frac12(1-\mathrm{i}) & \frac12(1+\mathrm{i}) & 0 \\
                           0 & 0 & 0 & 1
                         \end{array}}
                    \right).
$$
From Eq. (\ref{eq10}), we have
$$
\mathrm{\bf CGP_S}(\sqrt{\text{swap}}) =
\frac{1}{4}\left(1-\frac{54545\pi}{262144}\right)\approx 0.087.
$$

\begin{figure}[htbp]\centering
\includegraphics[width=0.5\textwidth]{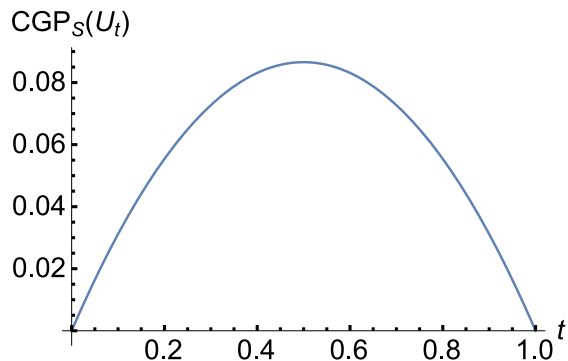}
\caption{The skew information-based coherence generating power of
$U_{t}$ with respect to the parameter $t$.} \label{fig:Fig3}
\end{figure}

{\bf Example 3.4} For a partial swap operator \cite{Audenaert2016},
one has $U_t\in \mathcal{U}(\mathbb{C}^d\otimes\mathbb{C}^d)$:
$U_t=\sqrt{t}\mathbb{I}_d\otimes\mathbb{I}_d+\mathrm{i}\sqrt{1-t}\,S$,
where $S=\sum^d_{i,j=1}|ij\rangle\langle ji|$ and $t\in[0,1]$. When
$d=2$, we have
$$
U_t = \left({\begin{array}{cccc}
            \sqrt{t}+\sqrt{1-t}\mathrm{i} & 0 & 0 & 0 \\
            0 & \sqrt{t} & \sqrt{1-t}\mathrm{i} & 0 \\
            0 & \sqrt{1-t}\mathrm{i} & \sqrt{t} & 0 \\
            0 & 0 & 0 & \sqrt{t}+\sqrt{1-t}\mathrm{i}
          \end{array}}
      \right).
$$
Then it follows from Eq. (\ref{eq10}) that
\begin{eqnarray}\label{eq11}
\mathrm{\bf CGP_S}(U_t) =
t(1-t)\left(1-\frac{54545\pi}{262144}\right),\quad t\in[0,1].
\end{eqnarray}
We plot $\mathrm{\bf CGP_S}(U_t)$ as the function of
$t\in[0,1]$ in Fig.~\ref{fig:Fig3}. It is found that the maximal
value of $\mathrm{\bf CGP_S}(U_t)$ is $\mathrm{\bf
CGP_S}(U_{\frac{1}{2}})=\frac14\left(1-\frac{54545\pi}{262144}\right)\approx
0.087$ attained at $t=\frac{1}{2}$.

It is pointed out that \cite{Zhang6} the possible values of the relative
entropy-based CGP $\mathrm{\bf CGP_R}$ form the closed interval $[0,\mathrm{ln}N-H_N+1]$,
where $H_N=\sum^N_{n=1}1/n$, and both $\mathrm{\bf CGP_R}(H)$ and the maximal value of $\mathrm{\bf CGP_R}(U_\theta)$ ($\theta\in [0,\pi]$) reach the maximal CGP of
qubit unitary channels, $\mathrm{ln}2-1/2\approx 0.193$. In
comparison, $\mathrm{\bf
CGP_R}(\sqrt{\text{swap}})=\frac{1}{2}\mathrm{ln}2\approx 0.347$ is
greater than the maximal CGP of unitary channels given by $4\times
4$ unitary matrices, $\mathrm{ln}4-H_4+1\approx 0.303$, while the
maximal value of $\mathrm{\bf CGP_R}(U_t)$, $t\in [0,1]$,
$\frac{1}{4}(2\mathrm{ln}2-1)\approx 0.097$, is less than it.

Note that similarly, for skew information-based CGP $\mathrm{\bf CGP_S}$, both
$\mathrm{\bf CGP_S}(H)$ and the maximal value of $\mathrm{\bf
CGP_S}(U_\theta)$ ($\theta\in [0,\pi]$) reach the maximal CGP of
unitary channels given by $2\times 2$ unitary matrices,
$\frac{1}{2}(1-\frac{3\pi}{16})\approx 0.205$. However, both
$\mathrm{\bf CGP_S}(\sqrt{\text{swap}})$ and the maximal value of
$\mathrm{\bf CGP_S}(U_t)$ for $t\in [0,1]$,
$\frac14\left(1-\frac{54545\pi}{262144}\right)\approx 0.087$, are
less than the maximal CGP of unitary channels given by $4\times 4$
unitary matrices,
$\frac{3}{4}\left(1-\frac{54545\pi}{262144}\right)\approx 0.26$.
Moreover, for the same unitary channel given in the above examples,
the skew information-based CGP $\mathrm{\bf CGP_S}$ are less than the relative
entropy-based CGP $\mathrm{\bf CGP_R}$ calculated in \cite{Zhang6}.

\vskip0.1in

\noindent {\bf 4. CGP as a random variable over the unitary group
under skew information-based coherence}

\vskip0.1in

We now regard $\mathrm{\bf CGP_S}(U)$ as a random variable over the
group of $N\times N$ unitary matrices $U(N)$ equipped with the Haar
measure $\mathrm{d\mu_{Haar}}(U)$. We calculate the mean value of
$\mathrm{\bf CGP_S}(U)$ for random unitary channels.

If $G$ is a locally compact group, there is, up to a constant
multiple, a unique regular Borel measure $\mu_L$ that is invariant
under left translation. Here left translation invariance of a
measure $\mu$ means that $\mu(M)=\mu(gM)$ for all measurable sets
$M$ and $g\in G$. Regularity means that
$\mu(M)=\inf\{\mu(\mathcal{O}): M\subseteq \mathcal{O},\
\mathcal{O}\ \makebox{open}\}=\sup\{\mu(\mathcal{C}): M\supseteq
\mathcal{C},\ \mathcal{C}\ \makebox{compact}\}$. Such a measure is
called a left-invariant Haar measure. It has the properties that any
compact set has finite measure and any nonempty open set has
positive measure. Left-invariance of the measure amounts to
left-invariance of the corresponding integral,
$$\int_{G}f(g'g)\mathrm{d}\mu_L(g)=\int_{G}f(g)\mathrm{d}\mu_L(g)$$ for any Haar integral
function $f$ on $G$ and any $g'\in G$ \cite{DBump}. We denote this
left-invariant Haar measure by $\mu_\mathrm{Haar}$ here.

{\bf Theorem 4.1} The mean value of $\mathrm{\bf CGP_S}(U)$ is given
by
\begin{equation}\label{eq12}
\mathbb{E}_{U}[\mathrm{\bf
CGP_S}(U)]=\frac{N-1}{N+1}\left(1-\frac{1}{N^2(N-1)}\left[\left(\sum_{k=1}^N
I_{kk}^{(\frac{1}{2})}\right)^2-\sum_{k,l=1}^N
\left(I_{kl}^{(\frac{1}{2})}\right)^2\right]\right).
\end{equation}
where
$I_{kl}^{(\frac{1}{2})}=\sum_{r=0}^{\min(k,l)}(-1)^{k+l}\tbinom{\frac{1}{2}}{k-r}\tbinom{\frac{1}{2}}{l-r}\frac{\Gamma(\frac{3}{2}+r)}{r!}
$.

{\bf Proof.} From Eq. (\ref{eq6}), the mean value of $\mathrm{\bf
CGP_S}(U)$ for unitary channels is given by
\begin{eqnarray}\label{eq13}
&&\mathbb{E}_{U}[\mathrm{\bf CGP_S}(U)]\nonumber\\
&&=\int_{U(N)}\mathrm{d\mu_{Haar}}(U)\left(1-\frac{1}{N^2(N-1)}\left[\left(\sum_{k=1}^N
I_{kk}^{(\frac{1}{2})}\right)^2-\sum_{k,l=1}^N
\left(I_{kl}^{(\frac{1}{2})}\right)^2\right]\right)\left(1-\frac{1}{N}\sum_{k,i=1}^N|U_{ki}|^4\right),~~~~
\end{eqnarray}
where $\mathrm{\mu_{Haar}}$ is a unitarily invariant uniform Haar
measure. Noting that the Haar measure is left-invariant, we obtain
\begin{eqnarray}\label{eq14}
&&\mathbb{E}_{U}[\mathrm{\bf CGP_S}(U)]\nonumber\\
&&=\left(1-\frac{1}{N^2(N-1)}\left[\left(\sum_{k=1}^N
I_{kk}^{(\frac{1}{2})}\right)^2-\sum_{k,l=1}^N
\left(I_{kl}^{(\frac{1}{2})}\right)^2\right]\right)\left(1-\frac{1}{N}\int_{U(N)}\mathrm{d\mu_{Haar}}(U)\sum_{k,i=1}^N|U_{ki}|^4\right)\nonumber\\
&&=\left(1-\frac{1}{N^2(N-1)}\left[\left(\sum_{k=1}^N
I_{kk}^{(\frac{1}{2})}\right)^2-\sum_{k,l=1}^N
\left(I_{kl}^{(\frac{1}{2})}\right)^2\right]\right)\left(1-N\int_{U(N)}\mathrm{d\mu_{Haar}}(U)|U_{11}|^4\right),~~
\end{eqnarray}
where $U_{11}=\langle 1|U|1\rangle$. From the proof of Theorem 1 in
\cite{Wuzhaoqi3}, we have
\begin{eqnarray}\label{eq15}
\int_{U(N)}\mathrm{d\mu_{Haar}}(U)|U_{11}|^4
&=&(N-1)B(3,N-1)=(N-1)\frac{\Gamma(3)\Gamma(N-1)}{\Gamma(N+2)}\nonumber\\
&=&(N-1)\frac{2}{(N+1)N(N-1)}=\frac{2}{N(N+1)}.
\end{eqnarray}
Substituting (\ref{eq15}) into (\ref{eq14}) one gets (\ref{eq12}).
$\Box$

Comparing (\ref{eq12}) in Theorem 4.1 with (\ref{eq19}) in Theorem 4
of \cite{Wuzhaoqi3}, we see that the mean value
$\mathbb{E}_{U}[\mathrm{\bf CGP_S}(U)]$ of skew information-based
CGP for random unitary channels coincides with average skew
information-based coherence $\mathbb{E}_{\rho}[C_I(\rho)]$ for
random quantum mixed states. This fact can be explained as follows.
On the one hand, any quantum mixed state can be diagonalized, i.e.,
$\rho=U\Lambda U^\dag$, where $\Lambda$ is a diagonal matrix and $U$
is a unitary operator. Hence, for a random quantum mixed state, the
randomness resides in both the diagonal matrix $\Lambda$ and the
unitary operator $U$. On the other hand, according to Eq.
(\ref{eq2}), the CGP of a unitary channel with respect to skew
information-based coherence is defined via probabilistic averages,
in which the probability measure is on a uniform ensemble of
incoherent states (diagonal matrices). Therefore, it is not
surprising that these two quantities are equal.

Define the normalized CGP $\widetilde{\mathrm{\bf
CGP_S}}(U):=\mathrm{\bf CGP_S}(U)/\mathrm{CGP}_N\leq 1$. Then it
follows that $\mathbb{E}_{U}[\widetilde{\mathrm{\bf
CGP_S}}(U)]=\frac{N}{N+1}$, which coincides with the normalized CGP
based on the Hilbert-Schmidt norm of coherence presented in
\cite{Zanardi1}. Let $(X,d_1)$ and $(Y,d_2)$ be two metric spaces
and $T:X\rightarrow Y$ be a mapping. $T$ is called a Lipschitz
continuous mapping on $X$ with the Lipschitz constant $\eta$, if
there exists $\eta>0$ such that $d_2(T(x),T(y))\leq \eta d_1(x,y)$
holds for all $x,y\in X$ \cite{MOSearcoid}.

Let $X:\mathrm{U(N)}\rightarrow \mathbb{R}$ be a Lipschitz
continuous function from the unitary group to the real line with a
Lipschitz constant $K$, i.e., $|X(U)-X(V)|\leq K\|U-V\|_2$, with
$\|\cdot\|_2$ denoting the Hilbert-Schmidt norm of a matrix $A$,
i.e., $\|A\|_2:=\sqrt{\mathrm{Tr}A^\dag A}$ \cite{MMWilde}. Let
$U\in \mathrm{U(N)}$ be chosen uniformly at random. Then for any
$\epsilon>0$, we have the Levy's Lemma for Haar-distributed $N\times
N$ unitaries \cite{GWAnderson}:
$\mathrm{Pr}\{|X(U)-\mathbb{E}[X(U)]|\geq \epsilon\}\leq
\mathrm{exp}\left(\frac{-N\epsilon^2}{4K^2}\right)$, where
$\mathrm{Pr}$ denotes the probability of a random event. Using this
version of Levy's Lemma, we can similarly obtain the typicality of
CGP, similar to the one presented in \cite{Zanardi1},
$$\mathrm{Pr}\left\{\widetilde{\mathrm{\bf
CGP_S}}(U)\geq 1-\frac{2}{N^{1/3}}\right\}\geq 1-
\mathrm{exp}\left(\frac{-N^{1/3}}{256}\right).$$

\vskip0.1in

\noindent {\bf 5. CGP of mixed unitary channels under skew
information-based coherence}

\vskip0.1in

In this section, we consider the convex combinations of unitary channels of
the form $\Phi(\cdot)=\sum_{m}p_mU_m \cdot U_{m}^\dag$. In this
case, $\sqrt{\Phi(\Lambda)}=\sqrt{\sum_{m}p_mU_m \Lambda
U_{m}^\dag}$. In general it is difficult to compute
$\mathrm{\bf CGP_S}(\Phi)$ since $\sqrt{\Phi(\Lambda)}$ is hard to
tackle. We present an upper bound for this class of channels.

{\bf Theorem 5.1} For mixed unitary channels
$\Phi(\cdot)=\sum_{m}p_mU_m \cdot U_{m}^\dag$,  we have
\begin{equation}\label{eq16}
\mathrm{\bf CGP_S}(\Phi)\leq \sum_{m}p_m\mathrm{\bf CGP_S}(U_m).
\end{equation}

{\bf Proof.} Note that the skew information-based coherence is a
well-defined coherence measure which satisfies the convexity under
classical mixing \cite{Yuchangshui1},
$$
C_S\left(\sum_n q_n\rho_n\right)\leq \sum_n q_nC_S(\rho_n),
$$
where $q_n\geq 0$ and $\sum_n q_n=1$. It follows that
$$
C_S(\Phi(\Lambda))=C_S\left(\sum_{m}p_mU_m \Lambda
U_{m}^\dag\right)\leq \sum_{m}p_m C_S(U_m \Lambda U_{m}^\dag).
$$
Therefore, by the definition (\ref{eq2}) we get
$$
\mathrm{\bf CGP_S}(\Phi)=\int_{\Gamma}\mathrm{d}\mu(\Lambda)C_S(\Phi(\Lambda))
\leq \sum_{m}p_m\int_{\Gamma}\mathrm{d}\mu(\Lambda)C_S(U_m \Lambda
U_{m}^\dag)=\sum_{m}p_m\mathrm{\bf CGP_S}(U_m).
$$
This completes the proof. $\Box$

As applications, we consider the Pauli channels defined by
\begin{equation}\label{eq17}
\Phi(\rho)=\sum_{m=0}^3 p_m\sigma_m\rho\sigma_m,~\,\,p_m\geq
0,\,\,~\sum_{m=0}^3p_m=1,
\end{equation}
where $\sigma_0=I$, and $\sigma_m$, $m=1,2,3$, are the standard
Pauli matrices. When $p_1=p_2=p_3=p$, one has the depolarizing
channel. When $p_1=p$, $p_2=p_3=0$ and $p_1=p_2=0$, $p_3=p$ one gets
the bit-flipping channel and the phase-flipping channel,
respectively. The case $p_2=p$ and $p_1=p_3=0$ corresponds to the
bit-phase-flipping channel. From Eq. (\ref{eq8}), it is easy to
check that $\mathrm{\bf CGP_S}(\sigma_m)=0$ for $m=0,1,2,3$. Hence,
for any Pauli channel $\Phi$ we have $\mathrm{\bf CGP_S}(\Phi)=0$
from (\ref{eq16}).

As another example, consider the (unital) amplitude damping channel $\Phi(\rho)=\sum_{n=1}^2
E_n\rho E_n^\dag$ with
\begin{equation*}
E_1=\left({\begin{array}{cc}
                             1 & 0 \\
                             0 & \sqrt{1-\gamma}
                             \end{array}}\right),\ \ \
E_2=\left({\begin{array}{cc}
                             0 & 0 \\
                             0 & \sqrt{\gamma}
                             \end{array}}\right).
\end{equation*}
We have $\Phi(\Lambda)=\Lambda$. It follows from (\ref{eq6}) that
$\mathrm{\bf CGP_S}(\Phi)=0$.

For the (nonunital) amplitude damping channel $\Phi(\rho)=\sum_{n=1}^2 E_n\rho E_n^\dag$ with
\begin{equation*}
E_1=\left({\begin{array}{cc}
                             1 & 0 \\
                             0 & \sqrt{1-\gamma}
                             \end{array}}\right),\ \ \
E_2=\left({\begin{array}{cc}
                             0 & 0 \\
                             \sqrt{\gamma} & 0
                             \end{array}}\right),
\end{equation*}
we have $$\Phi(\Lambda)=\left({\begin{array}{cc}
                             \lambda_1+\gamma \lambda_2 & 0 \\
                             0 & (1-\gamma)\lambda_2
                             \end{array}}\right),$$
where $\Lambda=\mathrm{diag}(\lambda_1,\lambda_2)$. This implies
that $C_S(\Phi(\Lambda))=0$, and thus $\mathrm{\bf CGP_S}(\Phi)=0$.

More generally, by the property (8) given in \cite{Luo3}, it can be seen that
if the Kraus operators $E_n$ of the channel $\Phi$ and the reference
basis $\{|k\rangle\}$ satisfy $E_n|k\rangle \langle k|=|k\rangle
\langle k|E_n$ for all $n$ and $k$, then we have
$C_S(\Phi(\rho))\leq C_S(\rho)$, and thus $C_S(\Phi(\Lambda))\leq
C_S(\Lambda)$. For such channels, it holds that $\mathrm{\bf
CGP_S}(\Phi)=0$.

\vskip0.1in

\noindent {\bf 6. Conclusions and discussions}

\vskip0.1in

We have introduced the measure of the coherence generating power (CGP) of a quantum
channel, which is the average coherence generated by the channel
acting on a uniform ensemble of incoherent states.
By adopting the technique of probabilistic averages,
we have successfully derived the explicit analytical
formulae of CGP for any arbitrary finite dimensional unitary channels.
Furthermore, we have formulated the mean value of the CGP over the
unitary group, and investigated the typicality of the normalized
CGP. Moreover, we have also presented an upper bound on CGP for mixed
unitary channels (the convex combination of unitary channels) by
using the convexity of the skew information-based coherence.
Detailed examples have been provided for quibt channels.

In \cite{Zanardi1,Zanardi2} Zanardi \emph{et al.} have studied the
coherence generating power of unitary channels based on the
coherence measure of Hilbert-Schmidt norm, in which the measure is
in fact not well-defined and the computation involves only integrals
in uniform Haar measure over pure states. Instead, the authors in
\cite{Zhang6} derived the formulae of CGP for unitary channels based
on the relative entropy of coherence. The skew information-based
coherence measure we adopted in this paper is also well-defined and
has many important operational interpretations.

The obtained results enrich and complement the ones given in
\cite{Zhang6}. It is also worth pointing out that as the dimension
$N\rightarrow \infty$, the CGP of unitary channels in
\cite{Zanardi1,Zanardi2} approaches to 0, while the CGP of unitary
channels in \cite{Zhang6} does not always approach to 0. The mean
value of the CGP for random unitary channels also approaches to 0
when $N\rightarrow \infty$. In comparison, numerical results show
that our CGP of unitary channels and the mean value of the CGP over
the unitary group both approaches to a positive number close to
0.28. Our results may shed some new light on the studies of
coherence generating power of quantum channels.

\vskip0.1in

\noindent

\subsubsection*{Acknowledgements}
The authors would like to express their sincere gratitude to the
anonymous referees for their comments and suggestions, which have
greatly improved this paper. This work was supported by National
Natural Science Foundation of China (Grant Nos. 12161056, 11971140,
12075159, 12171044, 11875034); Jiangxi Provincial Natural Science
Foundation (Grant No. 20202BAB201001); Beijing Natural Science
Foundation (Grant No. Z190005); Academy for Multidisciplinary
Studies, Capital Normal University; Shenzhen Institute for Quantum
Science and Engineering, Southern University of Science and
Technology (Grant No. SIQSE202001), the Academician Innovation
Platform of Hainan Province.

\subsubsection*{Competing interests}
The authors declare no competing interests.

\subsubsection*{Data availability}
Data sharing not applicable to this article as no datasets were
generated or analysed during the current study.

\vskip0.2in

{\bf Appendix A: Proof of Theorem 3.1}

{\bf Proof of Theorem 3.1} Suppose that the spectral decomposition
of $\Lambda$ is $\Lambda=\sum_{j=1}^N \lambda_{j}|j\rangle\langle
j|$. Then $\sqrt{\Lambda}=\sum_{j=1}^N
\sqrt{\lambda_{j}}|j\rangle\langle j|$. Consider the unitary channel
$\Phi_U(\Lambda)=U\Lambda U^{\dag}$. We have
\begin{eqnarray}\label{eq18}
\int_{\textcolor{blue}{\mathcal{I}}}\mathrm{d}\mu(\Lambda)C_S(\Phi_U(\Lambda))
&=&\int_{\textcolor{blue}{\mathcal{I}}}\mathrm{d}\mu(\Lambda)\left[1-\sum_{k=1}^N\langle
k|U\sqrt{\Lambda}U^{\dag}|k\rangle ^2\right] \nonumber\\
&=&\int_{\textcolor{blue}{\mathcal{I}}}\mathrm{d}\mu(\Lambda)-\sum_{k=1}^N\int_{\textcolor{blue}{\mathcal{I}}}\mathrm{d}\mu(\Lambda)\left\langle
k^{\otimes2}|U^{\otimes2}\sqrt{\Lambda}^{\otimes2}(U^{\otimes2})^{\dag})|k^{\otimes2}\right\rangle
\nonumber\\
&=&1-\sum_{k=1}^N\left\langle
k^{\otimes2}|U^{\otimes2}\int_{\textcolor{blue}{\mathcal{I}}}\mathrm{d}\mu(\Lambda)\sqrt{\Lambda}^{\otimes2}(U^{\otimes2})^{\dag})|k^{\otimes2}\right\rangle.
\end{eqnarray}

Since
\begin{eqnarray}\label{eq19}
&&\int_{\textcolor{blue}{\mathcal{I}}}\mathrm{d}\mu(\Lambda)\sqrt{\Lambda}^{\otimes2} \nonumber\\
&&=\int_{\textcolor{blue}{\mathcal{I}}}\mathrm{d\mu(\Lambda)}\sum_{1\leq
i=j\leq N}\sqrt{\lambda_i\lambda_j}|ij\rangle\langle
ij|+\int_{\textcolor{blue}{\mathcal{I}}}\mathrm{d\mu(\Lambda)}\sum_{1\leq
i\neq j\leq N}\sqrt{\lambda_i\lambda_j}|ij\rangle\langle
ij| \nonumber\\
&&=\int_{\textcolor{blue}{\mathcal{I}}}\mathrm{d\mu(\Lambda)}\sum_{i=1}^N\lambda_i|ii\rangle\langle
ii|+\int_{\textcolor{blue}{\mathcal{I}}}\mathrm{d\mu(\Lambda)}\sum_{1\leq
i\neq j\leq
N}\sqrt{\lambda_i\lambda_j}|ij\rangle\langle ij| \nonumber\\
&&=\frac{1}{N}\sum_{i=1}^N|ii\rangle\langle
ii|+C^{\mathrm{HS}}_N\int_{\mathbb{R}_+^N}\sum_{1\leq i\neq j\leq
N}\sqrt{\lambda_i\lambda_j}\delta\left(1-\sum_{j=1}^N\lambda_j\right)|\Delta(\lambda)|^2\prod_{j=1}^N
\mathrm{d}\lambda_j |ij\rangle\langle ij|,~~~~~~~
\end{eqnarray}
where $\Delta(\lambda)=\prod_{1\leq k<l\leq
N}(\lambda_l-\lambda_k)$, and $\int_{\mathbb{R}_+^N}
\sqrt{\lambda_i\lambda_j}\delta\left(1-\sum_{j=1}^N\lambda_j\right)|\Delta(\lambda)|^2\prod_{j=1}^N
\mathrm{d}\lambda_j $ are the same for $i\neq j$, we only need to
calculate
$$
\int_{\mathbb{R}_+^N}\sqrt{\lambda_1\lambda_2}\delta\left(1-\sum_{j=1}^N\lambda_j\right)|\Delta(\lambda)|^2\prod_{j=1}^N
\mathrm{d}\lambda_j.
$$

Denote
$$F(t)=\int_{\mathbb{R}_+^N}\sqrt{\lambda_1\lambda_2}\delta\left(t-\sum_{j=1}^N\lambda_j\right)|\Delta(\lambda)|^2\prod_{j=1}^N
\mathrm{d}\lambda_j.$$ By performing Laplace transform
$(t\rightarrow s)$ of $F(t)$, and letting $\mu_j=s\lambda_j,j=1,2$,
we get
\begin{eqnarray}\label{eq20}
\tilde{F}(s)&=&\int_{\mathbb{R}_+^N}\sqrt{\lambda_1\lambda_2}\mathrm{exp}\left(-s\sum_{j=1}^N
\lambda_j\right)|\Delta(\lambda)|^2\prod_{j=1}^N
\mathrm{d}\lambda_j  \nonumber\\
&=& s^{-(N^2+1)}\int_{\mathbb{R}_+^N}
\sqrt{\mu_1\mu_2}\mathrm{exp}\left(-\sum_{j=1}^N
\mu_j\right)|\Delta(\mu)|^2\prod_{j=1}^N \mathrm{d}\mu_j,
\end{eqnarray}
where $\Delta(\mu)=\prod_{1\leq k<l\leq N}(\mu_l-\mu_k)$. Utilizing
the inverse Laplace transform $(s\rightarrow
t):\,\mathscr{L}^{-1}(s^{\alpha})=\frac{t^{-\alpha-1}}{\Gamma(-\alpha)}$,
we obtain
\begin{equation}\label{eq21}
F(t)=\frac{t^{N^2}}{\Gamma(N^2+1)}\int_{\mathbb{R}_+^N}\sqrt{\mu_1\mu_2}\mathrm{exp}\left(-\sum_{j=1}^N
\mu_j\right)|\Delta(\mu)|^2\prod_{j=1}^N \mathrm{d}\mu_j.
\end{equation}
Thus
\begin{eqnarray}\label{eq22}
&&\int_{\mathbb{R}_+^N}\sqrt{\lambda_1\lambda_2}\delta\left(1-\sum_{j=1}^N\lambda_j\right)|\Delta(\lambda)|^2\prod_{j=1}^N
\mathrm{d}\lambda_j
\nonumber\\
&&=\frac{1}{\Gamma(N^2+1)}\int_{\mathbb{R}_+^N}\sqrt{\mu_1\mu_2}\mathrm{exp}\left(-\sum_{j=1}^N
\mu_j\right)|\Delta(\mu)|^2\prod_{j=1}^N \mathrm{d}\mu_j\nonumber\\
&&=\frac{(N-2)!\prod^{N}_{j=1}\Gamma(j)^2}{\Gamma(N^2+1)}\left[\left(\sum_{k=1}^N
I_{kk}^{(\frac{1}{2})}\right)^2-\sum_{k,l=1}^N
\left(I_{kl}^{(\frac{1}{2})}\right)^2\right],
\end{eqnarray}
where we have used the following result \cite{Wuzhaoqi3},
\begin{eqnarray}\label{eq23}
&&\int_{\mathbb{R}_+^N}\sqrt{\mu_1\mu_2}\,\mathrm{exp}\left(-\sum_{j=1}^N
\mu_j\right)|\Delta(\mu)|^2\prod_{j=1}^N \mathrm{d}\mu_j
\nonumber\\
&&=(N-2)!\prod^{N}_{j=1}\Gamma(j)^2\left[\left(\sum_{k=1}^N
I_{kk}^{(\frac{1}{2})}\right)^2-\sum_{k,l=1}^N
\left(I_{kl}^{(\frac{1}{2})}\right)^2\right],
\end{eqnarray}
with
$I_{kl}^{(\frac{1}{2})}=\sum_{r=0}^{\min(k,l)}(-1)^{k+l}\tbinom{\frac{1}{2}}{k-r}\tbinom{\frac{1}{2}}{l-r}\frac{\Gamma(\frac{3}{2}+r)}{r!}$.

From (\ref{eq6}), (\ref{eq21}) and (\ref{eq22}) we obtain
\begin{eqnarray}\label{eq24}
&&\int_{\textcolor{blue}{\mathcal{I}}}\mathrm{d}\mu(\Lambda)\sqrt{\Lambda}^{\otimes2}\nonumber\\
&&=\frac{1}{N}\sum_{i=1}^N|ii\rangle\langle
ii|+\frac{1}{N^3(N-1)}\left[\left(\sum_{k=1}^N
I_{kk}^{(\frac{1}{2})}\right)^2-\sum_{k,l=1}^N
\left(I_{kl}^{(\frac{1}{2})}\right)^2\right]\sum_{1\leq i\neq j\leq
N}|ij\rangle\langle ij|.
\end{eqnarray}
Combining (\ref{eq24}) and (\ref{eq23}), we get
\begin{eqnarray*}
&&\int_{\textcolor{blue}{\mathcal{I}}}\mathrm{d}\mu(\Lambda)C_S(\Phi_U(\Lambda))\nonumber\\
&&=1-\left(\frac{1}{N}\sum_{k,i=1}^N|U_{ki}|^4+\frac{1}{N^3(N-1)}\left[\left(\sum_{k=1}^N
I_{kk}^{(\frac{1}{2})}\right)^2-\sum_{k,l=1}^N
\left(I_{kl}^{(\frac{1}{2})}\right)^2\right]\sum_{k=1}^N\sum_{1\leq
i\neq j\leq
N}|U_{ki}|^2|U_{kj}|^2\right) \nonumber\\
&&=1-\frac{1}{N}\sum_{k,i=1}^N|U_{ki}|^4-\frac{1}{N^3(N-1)}\left[\left(\sum_{k=1}^N
I_{kk}^{(\frac{1}{2})}\right)^2-\sum_{k,l=1}^N
\left(I_{kl}^{(\frac{1}{2})}\right)^2\right]\left(N-\sum_{k,i=1}^N|U_{ki}|^4\right)\nonumber\\
&&=\left(1-\frac{1}{N^2(N-1)}\left[\left(\sum_{k=1}^N
I_{kk}^{(\frac{1}{2})}\right)^2-\sum_{k,l=1}^N
\left(I_{kl}^{(\frac{1}{2})}\right)^2\right]\right)\left(1-\frac{1}{N}\sum_{k,i=1}^N|U_{ki}|^4\right),
\end{eqnarray*}
which completes the proof. $\Box$

\end{document}